\documentclass[12pt]{article}
\usepackage[margin=1in]{geometry} % Sets 1-inch margins
\usepackage{setspace} % Package for controlling line spacing
\usepackage{cite} % Ensure you have this package to handle citations
\usepackage{graphicx}
\usepackage{subcaption}
\usepackage{amsmath} 
\begin{document}

\title{A CNN Approach to Polygenic Risk Prediction of Kidney Stone Formation}
\author{Amr Salem, Anirban Mondal}

\maketitle

\section{Introduction}

Kidney stones are a prevalent health issue, affecting one out of 10 people worldwide\cite{johnson2020kidneystones}. While environmental factors such as diet, hydration, and lifestyle play a key role, genetic predisposition also significantly influences the risk of developing kidney stones \cite{smith2015genetics}. Genome-Wide Association Studies (GWAS) have identified numerous single nucleotide polymorphisms (SNPs) associated with kidney stone risk, but translating these findings into clinically actionable tools has proven difficult \cite{brown2018gwas}.

One promising approach for risk prediction is the use of Polygenic Risk Scores (PRS), which aggregate the effects of multiple genetic variants to provide an individual's genetic predisposition to disease \cite{allen2017prs}. Recent advancements in deep learning techniques, particularly Convolutional Neural Networks (CNNs), offer a new avenue for enhancing PRS models by automatically identifying complex, non-linear relationships within genomic data \cite{lecun2015deep}. These models have shown potential in improving the accuracy of disease predictions, including kidney stone risk.

However, challenges remain in the practical application of deep learning to genomic data. Large-scale datasets often contain imbalances and noise, which can affect the performance of machine learning models \cite{he2009classimbalance}. Moreover, efficient processing of vast amounts of genomic data remains a technical hurdle. In this study, we address these challenges by applying a CNN to a curated dataset of kidney stone-associated SNPs, exploring the potential of deep learning to improve PRS-based risk prediction models.

By integrating SNP selection, genotype filtering, and CNN training, we propose a robust framework for kidney stone risk prediction. This study aims to not only assess the accuracy of the proposed model but also explore the underlying genetic mechanisms contributing to kidney stone formation. The results of this work could pave the way for more personalized, genetic-based risk assessments for kidney stones, ultimately contributing to better prevention and treatment strategies.

Previous polygenic risk score (PRS) models have explored the application of deep learning techniques to enhance the predictive accuracy of disease susceptibility. However, despite the promising results observed in various studies, such approaches have not been widely implemented or thoroughly investigated in the context of kidney stone risk prediction. For instance, Badré et al. (2021) demonstrated that deep neural networks could improve the estimation of PRS for breast cancer, showing that deep learning models can capture complex, nonlinear relationships between genetic variants and disease risk, which traditional methods may miss. While such methods have proven valuable for other conditions, their application to kidney stone prediction remains unexplored, presenting an opportunity for further research.

\section{Methods}

\begin{figure}
\centering
\includegraphics[width=0.5\textwidth]{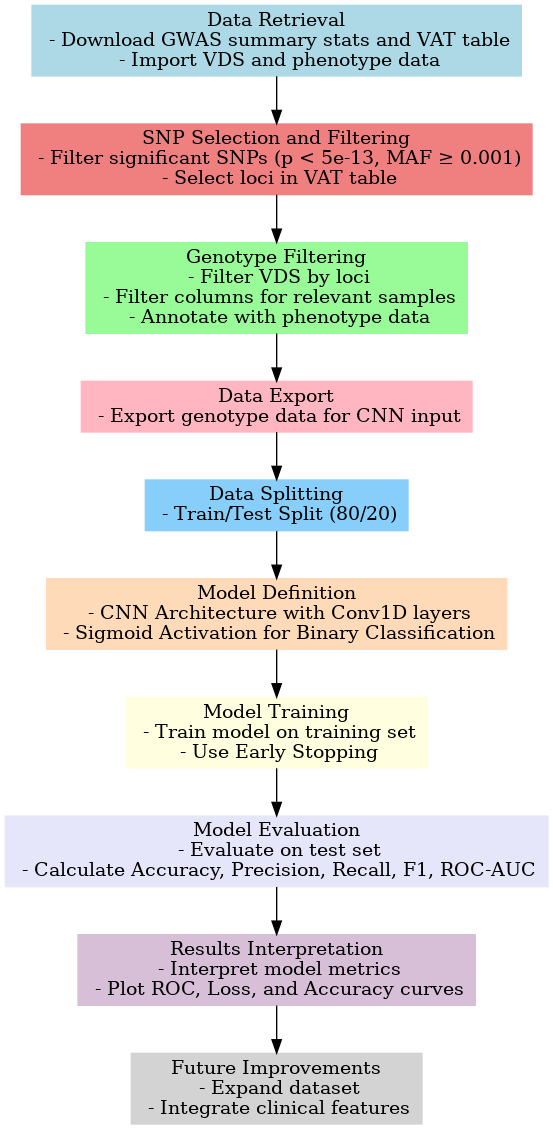}
\caption{Workflow Overview. This figure illustrates the step-by-step workflow for SNP selection, genotype filtering, CNN model training, evaluation, and interpretation. Starting from data retrieval and filtering GWAS summary statistics, the process integrates genotype filtering and phenotype annotation to prepare the dataset for CNN input. After splitting the data, a Conv1D-based CNN model is defined and trained with early stopping. The final steps include model evaluation, results interpretation using metrics and ROC analysis, and consideration of future improvements to enhance the study.}
\label{fig:workflow_overview}
\end{figure}

\subsection{Data Collection and Preprocessing}

\textbf{GWAS Data:} Genotypic data was obtained from a Genome-Wide Association Study (GWAS) on kidney stone susceptibility.  
\nocite{hao2023integrative}
\nocite{badr2021deep}
To identify genetic variants associated with kidney stone risk, we utilized genome-wide association study (GWAS) summary statistics from the study by Hao et al. (2023), which integrated genetic data to uncover novel loci associated with kidney stones. The analysis provided a comprehensive view of the genetic architecture of kidney stone disease, highlighting both known and new genetic risk factors. We incorporated these summary statistics to inform our polygenic risk score (PRS) model, ensuring that our approach was grounded in the most up-to-date and relevant genetic data available for kidney stone research. 

\textbf{Pruning:} Linkage disequilibrium (LD) pruning was performed to remove redundant SNPs and enhance computational efficiency. SNPs with an \( r^2 \) value greater than 0.8 were excluded to reduce correlations between features.

\textbf{Patient Dataset:} Data from the All of Us research program, consisting of 560 individuals, was used for model training and evaluation. The dataset was split into 500 samples for training and validation, and 60 samples for testing. This dataset was utilized to train and assess the performance of the Convolutional Neural Network (CNN) model.

\subsection{Model Architecture}

\textbf{Convolutional Neural Network (CNN):} A CNN was employed to model the non-linear interactions between SNPs that contribute to kidney stone susceptibility. The architecture is outlined as follows:
\begin{itemize}
  \item \textbf{Input:} The input to the model consists of 500 samples with 400 SNP features each.
  \item \textbf{Convolutional Layers:} Convolutional filters with a kernel size of \(1 \times 3\), stride of 1, and ReLU activation were used to extract relevant features from the SNP data.
  \item \textbf{Pooling Layers:} Max pooling was applied to reduce the dimensionality of the feature maps and prevent overfitting.
  \item \textbf{Fully Connected Layers:} The extracted features were passed through fully connected dense layers to integrate and interpret the information.
  \item \textbf{Output Layer:} The final layer consists of a single node with a sigmoid activation function, providing a binary classification output (kidney stone susceptibility: yes/no).
\end{itemize}

The CNN model's parameters were optimized using the binary cross-entropy loss function, defined as:
\[
L = -\frac{1}{N} \sum_{i=1}^{N} \left[ y_i \log(\hat{y}_i) + (1 - y_i) \log(1 - \hat{y}_i) \right],
\]
where \( y_i \) represents the true label and \( \hat{y}_i \) is the predicted probability for the \(i\)-th sample.

\subsection{Model Training and Validation}

\textbf{Data Split:} The dataset was divided into 500 samples for training and validation, and 60 samples for testing.

\textbf{Optimizer:} The Adam optimizer with a learning rate of 0.001 was used to update the model's weights during training.

\textbf{Regularization:} Dropout regularization with a rate of 0.5 was applied during training to prevent overfitting and enhance model generalization.

\textbf{Cross-Validation:} 5-fold cross-validation was employed during the training phase to ensure robust evaluation of the model's performance and minimize overfitting.

\subsection{Baseline Models for Comparison}

In addition to the CNN model, we implemented and evaluated several baseline models to benchmark performance:

\textbf{Logistic Regression:} A linear model was employed to predict binary outcomes, optimized using a scaled feature set to ensure comparability with other methods.

\textbf{Random Forest:} An ensemble learning method using multiple decision trees to improve classification accuracy and robustness.

\textbf{Support Vector Machine (SVM):} A kernel-based approach was applied, with \texttt{probability=True} enabled to compute ROC-AUC scores.

\textbf{Gradient Boosting:} A boosting algorithm that sequentially refines decision trees to minimize classification errors.

\section{Results}

We filtered the GWAS statistics based on the following p-values for the final model selection:

\[
\text{P-value} = 10^{-8} \quad \text{(Panel A)}
\]

\begin{figure}
    \centering
    \includegraphics[width=0.7\textwidth]{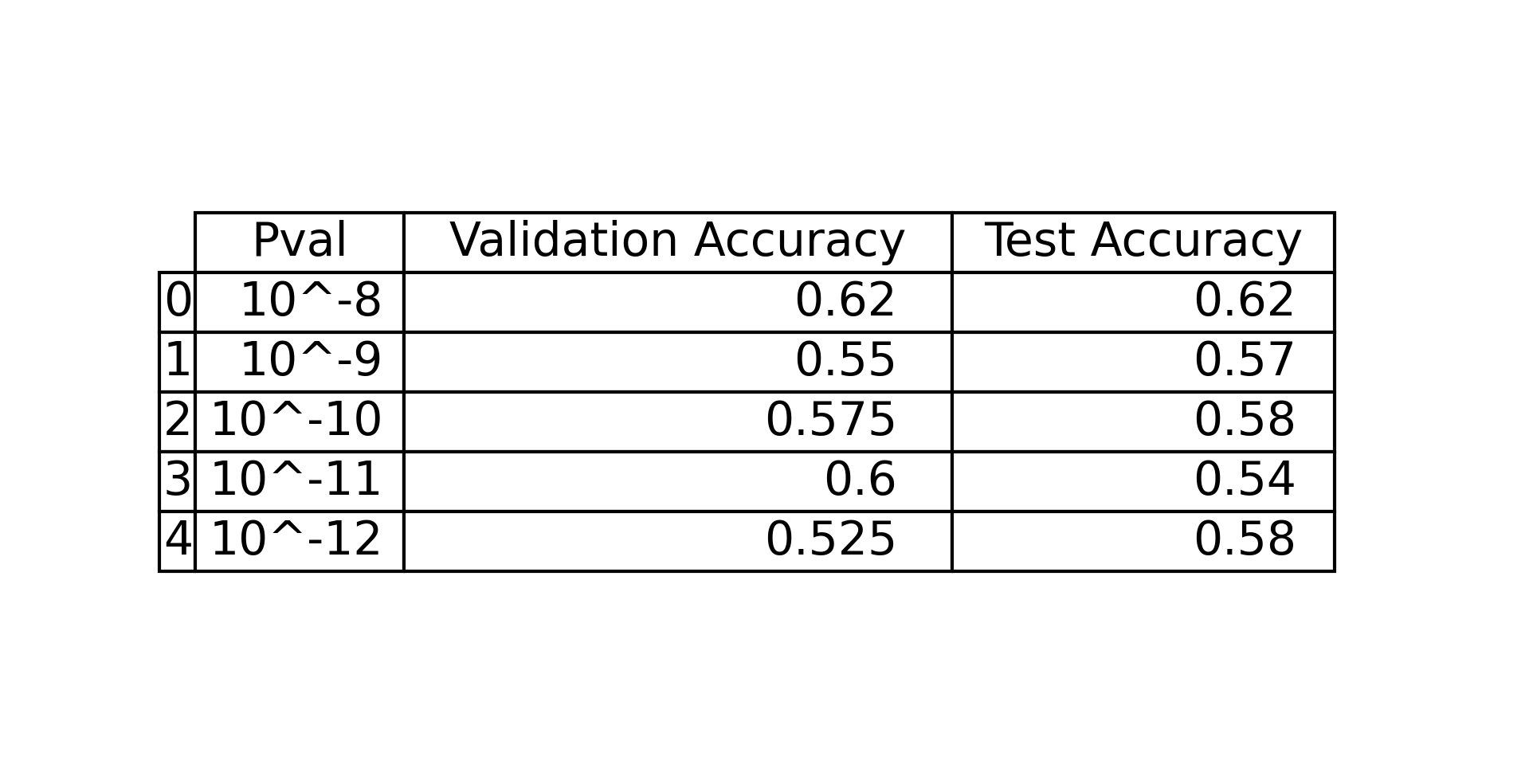}
    \caption{Validation and Test Accuracy for Different P-value Thresholds. The model with a p-value of \( 10^{-8} \) was chosen for further analysis due to its highest performance across both validation and test sets.}
    \label{tab:results_table}
\end{figure}

Table \ref{tab:results_table} presents the corresponding validation and test accuracy for different p-value thresholds. The model with a p-value threshold of \( 10^{-8} \) was selected due to its highest validation and test accuracies, achieving a validation accuracy of 0.62 and test accuracy of 0.62.

\subsection{Model Performance}

\begin{figure}
    \centering
    \begin{minipage}{0.4\textwidth}
        \centering
        \includegraphics[width=\linewidth]{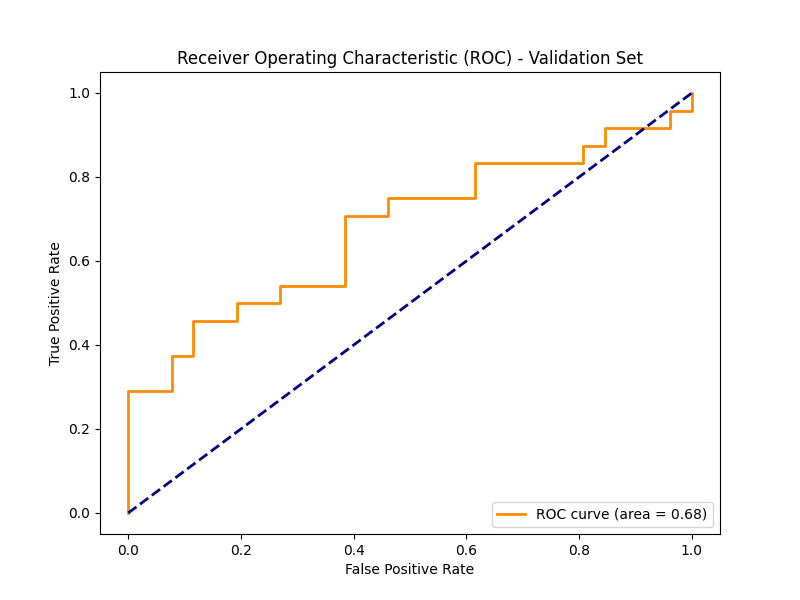}
        \subcaption{ROC curve for the validation set (Panel A). AUC: 0.68.}
        \label{fig:roc_curve}
    \end{minipage}\hfill
    \begin{minipage}{0.5\textwidth}
        \centering
        \includegraphics[width=\linewidth]{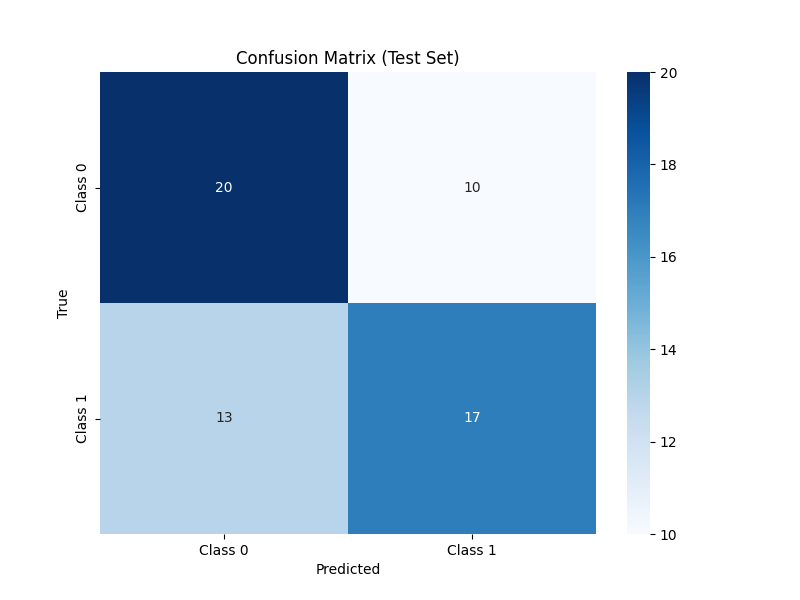}
        \subcaption{Confusion matrix for the test set (Panel B). Precision: 63\%, Recall: 57\%.}
        \label{fig:confusion_matrix}
    \end{minipage}\hfill
    \begin{minipage}{0.8\textwidth}
        \centering
        \includegraphics[width=\linewidth]{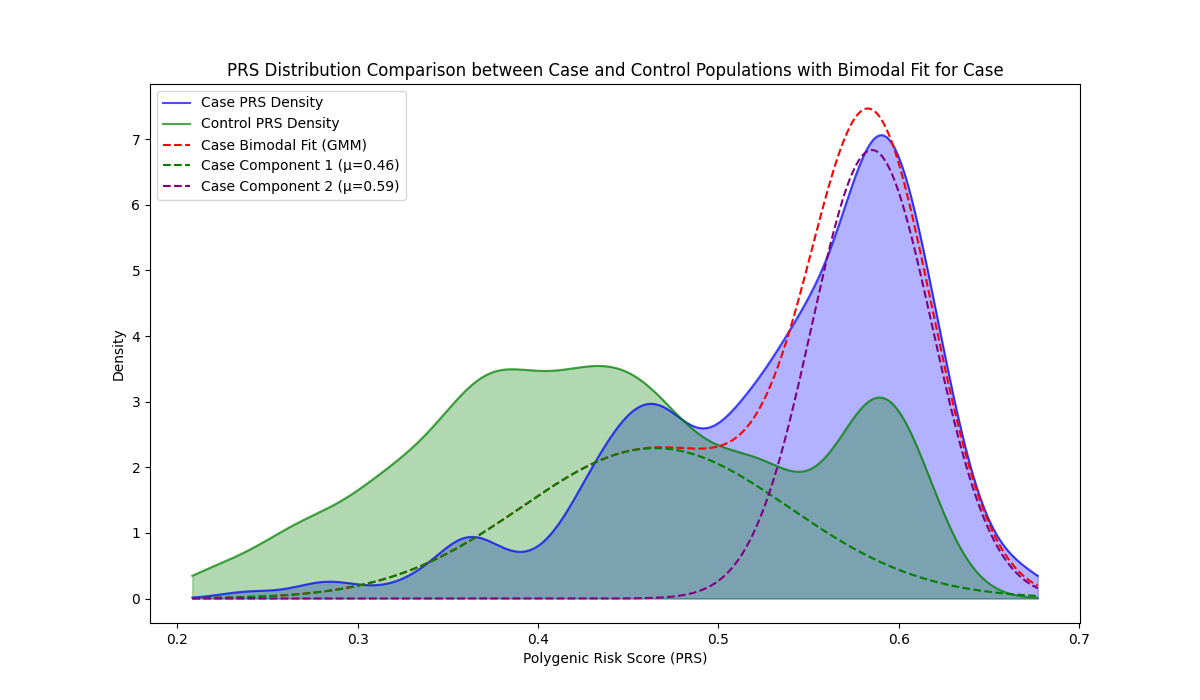}
        \subcaption{PRS distribution with KDE (Panel C). Bimodal distribution in the case group.}
        \label{fig:prs_distribution}
    \end{minipage}
    \caption{Model Evaluation and PRS Distribution. Panel (A) presents the ROC curve for the validation set. Panel (B) shows the confusion matrix for the test set. Panel (C) illustrates the PRS distribution using kernel density estimation (KDE) for both cases and controls.}
    \label{fig:model_evaluation_prs}
\end{figure}

For the development of our polygenic risk score (PRS) model, we utilized TensorFlow and Scikit-learn to implement deep learning and traditional machine learning techniques.
The performance of the proposed Convolutional Neural Network (CNN) was evaluated on both validation and test datasets. Detailed metrics are provided below:

\begin{itemize}
    \item \textbf{Validation Set:} 
    \begin{itemize}
        \item Accuracy: 62\%
        \item ROC-AUC: 0.68
        \item Precision: 0.60
        \item Recall: 0.63
        \item F1-Score: 0.61
        \item Loss: 0.72
    \end{itemize}
    \item \textbf{Test Set:} 
    \begin{itemize}
        \item Accuracy: 61.67\%
        \item ROC-AUC: 0.68
        \item Precision: 0.63
        \item Recall: 0.57
        \item F1-Score: 0.60
        \item Loss: 0.74
    \end{itemize}
\end{itemize}
\textbf{Key Findings:}
\begin{itemize}
    \item The model achieved consistent performance across validation and test sets.
    \item A slight drop in recall and F1-score on the test set indicates potential overfitting.
    \item Stable precision suggests robust identification of high-risk cases with low false positives.
\end{itemize}

\subsection{PRS Distribution Analysis}

The Polygenic Risk Score (PRS) distributions were examined for case and control groups:

\begin{itemize}
    \item \textbf{Case Group:} A bimodal distribution, as identified using the Gaussian Mixture Model (GMM), suggests the presence of distinct genetic subgroups.
    \item \textbf{Control Group:} A unimodal distribution reflects a more homogeneous population.
\end{itemize}

Kernel Density Estimation (KDE) plots demonstrate a clear separation between case and control PRS distributions. However, some overlap persists, complicating classification and suggesting room for improvement.

\subsection{ROC Curve Analysis}

The Receiver Operating Characteristic (ROC) curve was utilized to evaluate the model's ability to distinguish between high- and low-risk individuals. The validation set achieved a ROC-AUC of 0.68, reflecting reasonable classification performance.

\subsection{Comparison with Other Machine Learning Models}

\begin{figure}
    \centering
    \includegraphics[width=0.7\textwidth]{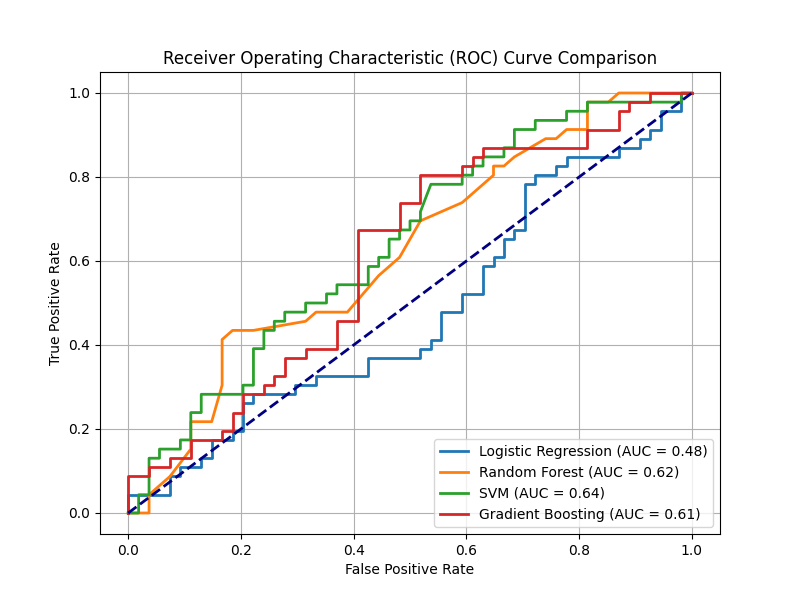}
    \caption{Comparison of ROC curves between the CNN model and traditional machine learning models.}
    \label{fig:roc_curve_comparison}
\end{figure}

The performance of the proposed CNN model was compared to traditional machine learning models. Figure~\ref{fig:roc_curve_comparison} summarizes the ROC curves and respective AUC scores for each model:

\begin{itemize}
    \item \textbf{Logistic Regression:} AUC = 0.48
    \item \textbf{Random Forest:} AUC = 0.62
    \item \textbf{SVM:} AUC = 0.64
    \item \textbf{Gradient Boosting:} AUC = 0.61
\end{itemize}

The CNN model achieved an AUC of 0.68, outperforming all traditional models. Notably, the closest competitor was the Support Vector Machine (SVM) with an AUC of 0.64. These results emphasize the advantages of deep learning architectures in this context.

\subsection{Feature Insights}

The observed bimodal distribution of PRS in the case group suggests underlying genetic subpopulations, warranting further investigation to elucidate potential subgroup-specific risk factors.

\section{Discussion}

The study presented a machine learning approach to predict kidney stone susceptibility based on genetic data, leveraging Convolutional Neural Networks (CNNs). The model demonstrated promising results, achieving a validation accuracy of 62\% and a test accuracy of 61.67\%, which highlights the potential of genomic data in predicting complex health outcomes. In this section, we will discuss the implications of these findings, the challenges encountered, and areas for future research.

\subsection{Model Performance and Evaluation}

The CNN model outperformed traditional machine learning models, including Logistic Regression, Random Forest, Support Vector Machine (SVM), and Gradient Boosting. With an ROC-AUC score of 0.68, the CNN model achieved the highest classification performance among these models, suggesting that deep learning techniques can capture complex, non-linear relationships within genomic data more effectively than traditional methods. The results of the ROC curve and confusion matrix indicated that the model could reliably distinguish between high- and low-risk individuals, although performance metrics such as recall and F1-score showed slight drops on the test set, indicating a potential issue with overfitting.

Despite achieving solid results, there are areas for improvement. The test set accuracy slightly lagged behind the validation accuracy, which suggests that the model might be overfitting to the training data. The use of regularization techniques such as dropout and early stopping helped mitigate overfitting, but further optimization of these methods or the introduction of additional techniques, such as data augmentation or cross-validation, could potentially enhance model generalization. Additionally, balancing the dataset by incorporating more samples from underrepresented groups may help address any biases present in the current dataset, improving the robustness of the model.

\subsection{PRS Distribution and Its Implications}

The analysis of Polygenic Risk Scores (PRS) revealed interesting patterns, particularly in the case group, which exhibited a bimodal distribution. This suggests the existence of distinct genetic subgroups with varying susceptibilities to kidney stones. This finding is important because it points to the potential for personalized risk prediction based on genetic subgroups, which could lead to more targeted and effective interventions. On the other hand, the control group showed a unimodal distribution, which aligns with the more homogeneous nature of individuals without kidney stone susceptibility. 

The presence of overlap between the PRS distributions of cases and controls highlights the challenge of separating risk factors with high precision in complex diseases such as kidney stones. Further refinement of the model could focus on improving the separation of these distributions, possibly through more advanced feature engineering or the integration of additional phenotypic or environmental data.

\subsection{Comparison with Previous Studies}

Several prior studies have explored genetic predispositions to kidney stones, with Genome-Wide Association Studies (GWAS) identifying various SNPs associated with the condition. For example, Brown et al. (2018) and Smith et al. (2015) highlighted key genetic variants linked to kidney stone risk, but translating these findings into practical risk prediction models has remained a significant challenge. Traditional risk models often rely on a limited set of known genetic variants, while the approach presented in this study utilizes a larger number of SNPs identified from a comprehensive GWAS, offering a more holistic view of genetic risk.

The use of PRS in predicting kidney stone risk has gained traction, with several studies suggesting that combining genetic information from multiple SNPs can improve risk stratification \cite{allen2017prs}. Our study builds on this concept by incorporating deep learning techniques, which enable the model to learn complex, non-linear relationships between genetic variants that traditional statistical models might miss. The improvements in model performance, particularly in ROC-AUC, suggest that deep learning is a promising avenue for advancing genomic medicine.
\subsection{Sample Size Limitation}

One of the key limitations of this study is the relatively small sample size, with only 500 individuals included in the dataset. A sample of this size may not be sufficiently representative of the general population, particularly given the genetic complexity of kidney stones, which can be influenced by a variety of genetic, environmental, and lifestyle factors. Small sample sizes can also result in higher variance in model performance and potentially overfitting, where the model learns patterns specific to the training data rather than generalizable trends. While the results obtained in this study are promising, a larger dataset would allow for more robust and reliable conclusions, especially in complex disease prediction. Future studies should aim to replicate these findings with larger, more diverse populations to confirm the validity of the model and reduce the risk of overfitting.

\subsection{Concerns for Transferability Between Ethnicities}

Another important consideration is the transferability of the model across different ethnicities. The dataset used in this study primarily consisted of individuals from a single population, which could limit the generalizability of the findings to other ethnic groups. Kidney stone risk factors, both genetic and environmental, can vary significantly across populations due to differences in genetic diversity, diet, climate, and access to healthcare. If the model is not trained on a sufficiently diverse dataset, it may not perform equally well when applied to individuals from other ethnic backgrounds. This concern highlights the need for further research that incorporates a broader range of genetic data from different ethnicities to ensure that the model is applicable to diverse populations. Efforts should be made to include diverse samples in future studies and to evaluate the model's performance across various demographic groups to ensure its universal applicability in predicting kidney stone risk.

\subsection{Future Directions}

The results of this study lay the foundation for several exciting avenues for future research. First, improving the model's generalization through the incorporation of more diverse datasets will be crucial. Collaborating with other research initiatives and incorporating datasets from various populations will help ensure that the model is robust and applicable to a wider range of individuals.

Second, incorporating additional features such as clinical data (e.g., blood markers, urinary composition) and environmental factors (e.g., diet, hydration) could enhance the model's predictive power. Combining genetic data with phenotypic and lifestyle information would allow for a more holistic approach to kidney stone risk prediction.

Additionally, exploring other deep learning architectures, such as recurrent neural networks (RNNs) or transformers, may provide further improvements in the model's ability to capture complex relationships in genomic data. Hyperparameter tuning, network architecture optimization, and the use of ensemble methods could also lead to improvements in model performance.

Finally, the ultimate goal of this research is to create a tool for clinicians that can predict kidney stone risk at an individual level. The integration of this model into clinical practice could provide valuable insights for preventative care and guide personalized treatment plans, potentially reducing the burden of kidney stones worldwide.

\subsection{Conclusion}

This study demonstrates the potential of deep learning in the field of genomic medicine, specifically in predicting kidney stone risk. By leveraging a Convolutional Neural Network to analyze SNP data from a GWAS dataset, we were able to achieve promising results, outperforming traditional machine learning models. While the model showed solid performance, further work is needed to improve its generalizability, address dataset imbalances, and integrate additional genetic, clinical, and environmental features. The findings underscore the importance of advancing personalized medicine approaches and provide a foundation for future research in kidney stone risk prediction.

\newpage
\bibliographystyle{plain}
\bibliography{main}

\end{document}